\documentclass[showpacs,reprint,twocolumn]{revtex4-1}

\usepackage{amsmath}
\usepackage{graphicx,xcolor}	
\usepackage{pstricks}		
\usepackage{pst-node}		

\newcommand{\eq}[1]{\begin{equation}{#1}\end{equation}} 

\newcommand{\figref}[1]{Fig.~\ref{#1}}		
\newcommand{\secref}[1]{Sec.~\ref{#1}}		

\newcommand{\diff}[2]{\frac{{\rm d} {#1}}{{\rm d{#2}}}} 
\newcommand{\vect}[1]{{#1}}			   
\newcommand{\matr}[1]{{\rm \bf {#1}}}		   

\newcommand{\OrbC}[1]{\matr{C}^{#1}}		
\newcommand{\ttfrac}[2]{\tfrac{#1}{#2}} 	

\newrgbcolor{subo1}{0.97 0.97 0} 	
\newrgbcolor{subo2}{0.8 0.8 0} 		
\newrgbcolor{orb1}{0.95 0 0}    	
\newrgbcolor{orb2}{0 0.92 0.92}    	
\newgray{attorb}{0.85}     		

\begin{document}
 
\title{Mesoscale symmetries explain dynamical equivalence of food webs}

\author{Helge \surname{Aufderheide}}
\author{Lars \surname{Rudolf}}
\author{Thilo \surname{Gross}}
\affiliation{University of Bristol, Merchant Venturers School of Engineering, Bristol, UK}
\date{\today}

\begin{abstract}
A present challenge in complex systems is to identify mesoscale structures that have distinct dynamical implications. In this paper we present a detailed investigation of a previously observed dynamical equivalence of certian ecological food webs. We show that this equivalence is rooted in mesoscale symmetries that exist in these webs. Certain eigenvectors of the Jacobian describing dynamical modes of the system, such as specific instabilities or responses to perturbations, localize on these symmetric motifs. On the one hand this means that by removing a symmetry from the network one obtains a system which has identical dynamics except for the removal of the localized mode. This explains the previously observed equivalence. On the other hand it means that we can identify dynamical modes that only depend on the symmetric motif. Symmetric structures thus provide an example for mesoscale network motifs having distinct and exact implications for the dynamics.
\end{abstract}


\maketitle 

\section{Introduction}
\label{sec:Introduction}
Making progress in the investigation of complex systems requires finding concepts that allow reducing the complexity, while leaving emergent-level features of the system intact. Networks offer a framework facilitating this reduction \cite{Newman2003,Dorogovtsev2002,Boccaletti2006,AlbertBarabasi2002}.
By representing the system as a set of discrete nodes connected by links, networks neglect much of the internal complexity of the system's constituents, but retain the complexity of their interaction.
Thereby, they offer a suitable intermediate level of description on which a qualitative understanding of the systems dynamics can be gained.

In the present paper we consider specifically the example of ecological food webs \cite{Ings2009}, i.e. the networks of \emph{who-eats-who} in ecology.
In these networks, the nodes represent distinct populations, whereas the links represent predator-prey interactions.
Food webs offer a highly simplified description of ecological systems, neglecting for instance non-predatory interactions and intra-population dynamics.
Yet, they capture the complexity of the network of predatory interactions that forms the backbone of most natural ecosystems.

A detailed analysis of food web dynamics, typically comprising 50-500 populations on very different time scales, poses a considerable challenge.
Therefore, one often considers coarse-grained models in which different populations are grouped and represented by one aggregate variable.
A classical example are food chain models in which whole layers of the food web such as phytoplankton, zooplankton, or fish are described by a single variable, respectively \cite{Steele1992,Iwasa1987}.
More biological information is preserved by aggregating so-called \emph{functional groups} or \emph{guilds} comprising species which are similar in morphology, life strategy, or niche \cite{Adams1985,Iwasa1987}.
Alternatively, it has been proposed to aggregate those species that hold similar topological positions in the network \cite{Solow1998,Polis1996,Girvan2002,Newman2006}.

From a theoretical perspective, the problem of coarse-graining networks is interesting because it touches upon mesoscale network properties.
We seek to identify a set of nodes that are sufficiently similar, such that their aggregation induces only minor changes in the dynamics.
The mesoscale properties of food webs have been previously studied in terms of community detection \cite{Girvan2002,Newman2006,Fortunato2010} and motif distributions \cite{Milo2002,Alon2007,Bascompte2005b}.
Although both communities and motifs are known to have a strong impact on the dynamics of the system, the precise nature of this impact is not known.

The ultimate goal of coarse-graining is finding a reduced representation of the network that conserves the dynamics of the system.
This goal is generally hard to achieve as it seems to imply knowledge of the dynamics of the original system.
However, in a recent publication \cite{ThiloFeudel2008} it was shown in a class of generalized models \cite{Thilo2006,ThiloLars2009} that under certain conditions an aggregation is possible that conserves the local dynamics around steady states and at least certain features of the global dynamics exactly.
Although the previous paper demonstrated this aggregation in several examples, it failed to provide a detailed analysis of the underlying symmetries that makes the aggregation possible.

In the present paper we build on recent advances \cite{MacArthur2009} in the dynamical implications of \emph{graph orbits}, which capture mesoscale symmetries in the network structure and extend them to directed and weighted networks.
Application to food web dynamics reveals the topological mechanism behind the reduction rule identified heuristically in Ref.~\onlinecite{ThiloFeudel2008} and provides more general rules for food web aggregation.
We note that the previously observed identity is of limited use for the analysis of food web data. 
However, its theoretical investigation provides an example of mesoscale structures having distinct implications on the network dynamics. 

The paper starts, in \secref{sec:previous}, with a brief review of previous results. 
In \secref{sec:Seperation} we introduce the notion of symmetries in food webs. Furthermore, we show that the nodes of a symmetry carry localized dynamical modes that characterized solely by the mesoscale structure of their symmetry.
In \secref{sec:Results} we establish the dynamical modes for several example food web symmetries and derive the coarse-graining rules allowing to remove dynamical modes of on the mesoscale.

\section{Dynamical Equivalence of Food Webs\label{sec:previous}}
The application considered below builds on previous results that have been obtained with a generalized model (GM) of ecological food webs \cite{Thilo2006,ThiloLars2009}. In this section we provide an overview of the approach of generalized modelling and the previous results on food web stability.

A GM is a system of differential equations in which not all processes are restricted to specific functional forms \cite{Thilo2006}.
For instance, the food web model studied below describes the dynamics of $N$ populations $X_1,\ldots,X_N$ by $N$ differential equation of the form
\begin{equation}
\diff{}{t}X_i  = G_i(\vect{X}) + S_i(X_i) - L_i(\vect{X}) -M_i(X_i),
\end{equation}
where $G_i$, $L_i$, $M_i$, and $S_i$ are unspecified functions describing the gain by predation ($G_i$), the loss by the predation($L_i$), the loss due to mortality ($M_i$), and the gain by reproduction ($S_i$).

Generalized models are typically too general to compute the number or location of steady states without further assumptions.
Yet, they can be analysed by local stability analysis.
For this purpose, we assume that a steady state $\vect{X}^* = (X_1^*,\ldots,X_N^*)^T$ were known and then formally linearize the system around this unspecified steady state.
The coefficients of the linearized system can be expressed by the Jacobian matrix, $J_{ij}=\ttfrac{\partial}{\partial X_j} (\ttfrac{{\rm d} }{{\rm dt}}X_i )|_*$, which depends on the derivatives of the unspecified functions in the model.
For instance, in the food web model this unspecified information appears in expressions such as
$\left.\ttfrac{\partial G_i}{\partial X_i}\right|_*$.
This derivative describes an unknown but constant quantity or, in other words, an unknown scalar parameter of the system.
Because this parameter is hard to interpret, we use a slightly different parametrization, which is obtained either by a special normalization procedure \cite{Thilo2006} or directly by the identity
\begin{equation}
\label{eq:Identity}
\left.\frac{\partial G_i}{\partial X_i}\right|_*=\frac{G_i^*}{X_i^*} \left.\frac{\partial \log G_i}{\partial \log X_i}\right|_*,
\end{equation}
which is true for $G_i^*,X_i^*>0$ (a condition that is generally met by definition; the special case of $X_i^*=0$ is discussed in Ref.~\onlinecite{Kuehn2010}).

The expression on the right-hand-side of \eqref{eq:Identity} is a product of two factors that can be interpreted directly in most applications.
The first factor is a per-capita rate.
Such rates have the dimension of inverse time and can be directly interpreted as characteristic turnover rates, i.e.~the inverse of the mean life expectancy of individuals.
The second factor is a logarithmic derivative.
Such derivatives are also called elasticities and have been proposed originally in economic theory \cite{Houthakker1969} and are also used in the context of metabolic control theory \cite{Fell1992}. 
They can be estimated well from data and interpreted straightforwardly.
For every power-law, $f(x)=Ax^p$, the logarithmic derivative is $\ttfrac{\partial \log f}{\partial \log p} = p$, independently of $A$ or $x$.
Thus, for instance, any linear function has an elasticity of one regardless of the slope.
For functions that are not power-laws the elasticity still provides an intuitive nonlinear measure of the sensitivity in the steady state.
\begin{figure}[tb]
\centering
\includegraphics[width=0.5\textwidth]{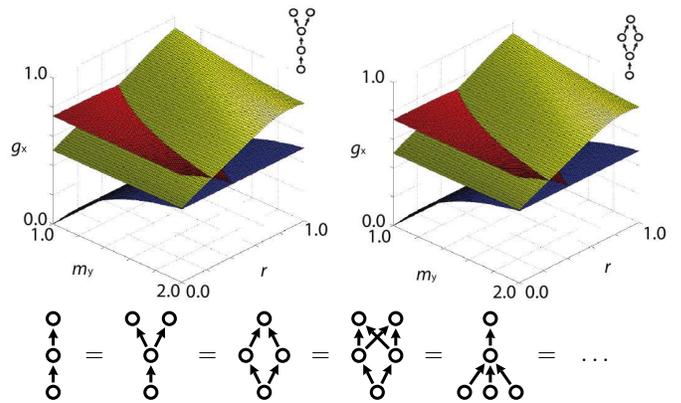}
\\
\def\chain{\begin{pspicture}(0.4,-0.15)(0.6,2.15)
\cnode(0.5,0){\radius}{C1}
\cnode(0.5,1){\radius}{C2}
\cnode(0.5,2){\radius}{C3}
\ncline{->}{C1}{C2}
\ncline{->}{C2}{C3}
\end{pspicture}}

\def\OOT{\begin{pspicture}(0.4,-0.15)(1.6,2.15)
\cnode(1,0){\radius}{C1}
\cnode(1,1){\radius}{C2}
\cnode(1.5,2){\radius}{C3}
\cnode(0.5,2){\radius}{D3}
\ncline{->}{C1}{C2}
\ncline{->}{C2}{C3}
\ncline{->}{C2}{D3}
\end{pspicture}}

\def\OTO{\begin{pspicture}(0.4,-0.15)(1.6,2.15)
\cnode(1,0){\radius}{C1}
\cnode(1.5,1){\radius}{C2}
\cnode(1,2){\radius}{C3}
\cnode(0.5,1){\radius}{D2}
\ncline{->}{C1}{C2}
\ncline{->}{C2}{C3}
\ncline{->}{C1}{D2}
\ncline{->}{D2}{C3}
\end{pspicture}}

\def\OTT{\begin{pspicture}(0.4,-0.15)(1.6,2.15)
\cnode(1,0){\radius}{C1}
\cnode(1.5,1){\radius}{C2}
\cnode(0.5,2){\radius}{C3}
\cnode(0.5,1){\radius}{D2}
\cnode(1.5,2){\radius}{D3}
\ncline{->}{C1}{C2}
\ncline{->}{C1}{D2}
\ncline{->}{C2}{C3}
\ncline{->}{D2}{C3}
\ncline{->}{C2}{D3}
\ncline{->}{D2}{D3}
\end{pspicture}}

\def\HOO{\begin{pspicture}(0.4,-0.15)(1.6,2.15)
\cnode(1,0){\radius}{C1}
\cnode(1,1){\radius}{C2}
\cnode(1,2){\radius}{C3}
\cnode(0.3,0){\radius}{D1}
\cnode(1.7,0){\radius}{E1}
\ncline{->}{C1}{C2}
\ncline{->}{C2}{C3}
\ncline{->}{D1}{C2}
\ncline{->}{E1}{C2}
\end{pspicture}}
 \psset{yunit=0.49,xunit=0.49,fillstyle=solid,linewidth=0.04,linecolor=black,fillcolor=white, nodesep=0.03, arrowinset=0.1, arrowlength=1}
 \def\radius{0.1}
\begin{tabular}{c@{}c@{}c@{}c@{}c@{}c@{}c@{}c@{}c@{}c@{}c}
  \parbox[c]{0.2cm}{\chain} & $\quad=\quad$ & \parbox[c]{0.6cm}{\OOT} & $\quad=\quad$ &\parbox[c]{0.6cm}{\OTO} & $\quad=\quad$ & \parbox[c]{0.6cm}{\OTT} & $\quad=\quad$ & \parbox[c]{0.6cm}{\HOO} & $\quad=\quad$ & $\ldots$
\end{tabular}
\caption{Equivalence of the bifurcation diagrams of the dynamics in food webs. 
The bifurcation diagrams of different food web models are exactly identical, if plotted as a function of the generalized parameters, as shown in the three-parameter bifurcation diagrams in the top row (reprinted from Ref.~\onlinecite{ThiloFeudel2008}). 
The parameter space in this example is spanned by three generalized parameters denoting the nonlinearity of mortality $m$, the sensitivity of predators to prey abundance $g$, and the ratio between the timescales of predator and prey dynamics $r$.
Every point in the parameter space corresponds to a specific steady state. 
The surfaces in the diagrams mark the location of bifurcation points of Hopf (red, green) and saddle-node (blue) bifurcations.
Steady states are stable if they lie in the front volume of the bifurcation diagram.
The two diagrams correspond to different food webs, indicated symbolically in the corner. 
In the symbolic representation nodes indicate populations and arrows indicate biomass flows due to predator-prey interactions. 
Using this notation the dynamical equivalence conjectured in Ref.~\onlinecite{ThiloFeudel2008} is shown symbolically in the lower row.}
\label{fig:equivDiag}
\end{figure}

From the parametrized Jacobian the dynamics close to the steady state under consideration can be computed. The steady state is stable if all eigenvalues of the Jacobian have negative real parts \cite{Kuznetsov1998}. Stability is lost when a change of parameters causes at least one eigenvalue to cross the imaginary axis, leading to a local bifurcation. Because the Jacobian is a real matrix the eigenvalues in question can be either real or form a complex conjugate eigenvalue pair. In the former case a bifurcation of saddle-node type occurs in which the number of steady states changes, whereas in the latter case a Hopf bifurcation occurs which marks the onset of (possibly transient) oscillations.  

The mathematical validity of generalized modelling was recently confirmed by rigorous proofs \cite{Kuehn2010}, whereas applicability to actual problems from different fields was demonstrated in several recent publications \cite{ThiloLars2009,Zumsande2010,Zumsande2010b,Stiefs2010,Gehrmann2010,Reznik2010,Yeakel2010,Gehrmann2011}.

A GM for ecological food webs \cite{Thilo2006} was used  in Ref.~\onlinecite{ThiloFeudel2008} to study critical points in parameter space in which the dynamics around steady states changes in bifurcations. The authors noted that in the context of the GM food webs with different topologies and different numbers of species  may be described by exactly identical local bifurcation diagrams when these are drawn in the generalized parameter space.
An example of the observed identity is reprinted in \figref{fig:equivDiag}. The figure shows three-parameter bifurcation diagrams for two different food webs for which the surfaces formed by bifurcation points of codimension-one bifurcations are identical.

The main conclusion of the previous paper \cite{ThiloFeudel2008} can be summarized in the conjecture that two species that a) interact with the same set of topological neighbours and b) are described by the same generalized parameters, can be aggregated into a single variable without changing the local bifurcation points. 

We note that the conjectured identity of bifurcation diagrams is not trivially true because of the nonlinearity of the underlying equations. 
Indeed, the identity is much harder to observe in conventional models and was hence not noted before the publication of Ref.~\onlinecite{ThiloFeudel2008}. 

Finally, consider that the identity extends also to local bifurcations of higher codimension (found for instance at the intersections of bifurcation surfaces), which have implications for the global dynamics of the system \cite{Thilo2005,Kuznetsov1998}. So the proposed aggregation must also conserve at least some features of the global dynamics.

\section{Mathematical treatment of symmetries \label{sec:Seperation}}
The equivalence of food webs described above is contingent on the presence of a specific symmetry in the topology of the food web.
In particular, equivalent reduction of the network was found to be possible if at least two populations
in the network interacted with exactly the same neighbours.

In this section we review and extend some previous results on the implications of symmetries in networks. 
Furthermore, we introduce the terminology that is used below for establishing a formal link between mesoscale symmetries and dynamics. 
We note that our presentation follows in large parts Ref.~\onlinecite{MacArthur2009} but extends the theory presented therein to directed networks. 

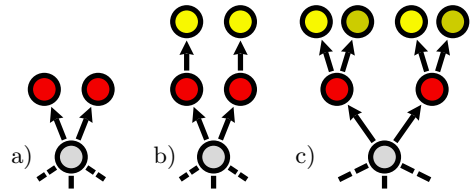
\begin{figure}[tb]
\centering
 \psset{yunit=0.9,xunit=1.1,fillstyle=solid,linewidth=0.07,linecolor=black,fillcolor=white, nodesep=0.03, arrowinset=0.1, arrowlength=1}
\def\radius{0.22}
\def\dorbit{\begin{pspicture}(-0.15,-0.5)(1.15,2.2)
\rput[l]{0}(-0.15,0){b)}
\cnode[fillcolor=subo1](0.25,2){\radius}{D1}
\cnode[fillcolor=subo1](0.9,2){\radius}{D2}
\cnode[fillcolor=orb1](0.25,1){\radius}{E1}
\cnode[fillcolor=orb1](0.9,1){\radius}{E2}
\cnode[fillcolor=attorb](0.575,0){\radius}{A}

\pnode(0.15,-0.35){W1}
\pnode(0.575,-0.5){W2}
\pnode(1,-0.35){W3}
\ncline[linestyle=dashed,dash=0.15 0.05]{A}{W1}
\ncline[linestyle=dashed,dash=0.15 0.05]{A}{W2}
\ncline[linestyle=dashed,dash=0.15 0.05]{A}{W3}

\ncline{->}{E1}{D1}
\ncline{->}{E2}{D2}
\ncline{->}{A}{E2}
\ncline{->}{A}{E1}

\end{pspicture}}
\def\corbit{\begin{pspicture}(-0.15,-0.5)(1.15,2.2)
\rput[l]{0}(-0.15,0){a)}
\cnode[fillcolor=orb1](0.25,1){\radius}{E1}
\cnode[fillcolor=orb1](0.9,1){\radius}{E2}
\cnode[fillcolor=attorb](0.575,0){\radius}{A}

\pnode(0.15,-0.35){W1}
\pnode(0.575,-0.5){W2}
\pnode(1,-0.35){W3}
\ncline[linestyle=dashed,dash=0.15 0.05]{W1}{A}
\ncline[linestyle=dashed,dash=0.15 0.05]{W2}{A}
\ncline[linestyle=dashed,dash=0.15 0.05]{W3}{A}

\ncline{->}{A}{E2}
\ncline{->}{A}{E1}

\end{pspicture}}
\def\suborbit{\begin{pspicture}(0,-0.5)(2.15,2.2)
\rput[l]{0}(0,0){c)}
\cnode[fillcolor=subo1](0.25,2){\radius}{D1}
\cnode[fillcolor=subo1](1.4,2){\radius}{D2}
\cnode[fillcolor=subo2](0.75,2){\radius}{D0}
\cnode[fillcolor=subo2](1.9,2){\radius}{D3}

\cnode[fillcolor=orb1](0.5,1){\radius}{E1}
\cnode[fillcolor=orb1](1.65,1){\radius}{E2}
\cnode[fillcolor=attorb](1.075,0){\radius}{A}

\pnode(0.5,-0.35){W1}
\pnode(1.075,-0.5){W2}
\pnode(1.65,-0.35){W3}
\ncline[linestyle=dashed,dash=0.15 0.05]{W1}{A}
\ncline[linestyle=dashed,dash=0.15 0.05]{W2}{A}
\ncline[linestyle=dashed,dash=0.15 0.05]{W3}{A}

\ncline{->}{E1}{D0}
\ncline{->}{E2}{D3}
\ncline{->}{E1}{D1}
\ncline{->}{E2}{D2}
\ncline{->}{A}{E2}
\ncline{->}{A}{E1}

\end{pspicture}}
\corbit
\hspace{10pt}
\dorbit
\hspace{10pt}
\suborbit
\caption{Three examples of symmetric structures in networks. Orbits are marked by the colouring of the corresponding nodes.  
Simple symmetric structures can consist of a single orbit (a, red), or multiple orbits (b, red, yellow), and can potentially contain nested orbits (c, red, yellow). See text for details. 
}
\label{fig:orbitfig}
\end{figure}

We say that a network contains a mesoscale \emph{symmetry} if there is a permutation of nodes that leaves the adjacency relations unchanged. 
For instance \figref{fig:orbitfig}a) shows a small part of a larger network in which two nodes can be exchanged without changing the adjacency. 
We note that if multiple symmetries exist they can be combined into larger symmetries, but for simplicity we assume that the symmetries are \emph{minimal}, i.e. that they cannot be decomposed into symmetries acting on fewer nodes.

In the following we denote all nodes that participate in a given minimal symmetry as a \emph{symmetric structure}. 
Further, a set of nodes that are mapped onto each other under a minimal symmetry operation are said to form an \emph{orbit}.

In the simplest case a symmetric structure contains only one such orbit (\figref{fig:orbitfig}a).
However, some symmetric structures comprise multiple orbits such that multiple permutations have to be carried out in parallel on different sets of nodes to preserve adjacency (\figref{fig:orbitfig}b).

We refer to the number of nodes belonging to an orbit as the \emph{size} of the orbit.
The size corresponds to the number of identical graphs generated by the corresponding symmetry operation unless a symmetric structure contains nested orbits (\figref{fig:orbitfig}c).

Although the concept of orbits was originally proposed for unweighted undirected networks, it can be intuitively extended to weighted and directed ones. In this case we additionally demand that symmetries conserve link direction and weight, such that the 
corresponding permutation leaves the matrix of link weights of the network unchanged.  

In Ref.~\onlinecite{MacArthur2009} it is shown that  symmetries in undirected networks have a distinct impact on the spectrum of the adjacency matrix.
In particular, if a symmetric structure is present then the eigenvalues can be decomposed into two classes. The first class contains the so-called \emph{redundant eigenvalues}. These eigenvalues correspond to eigenvectors that are localized on the symmetric structure. The second class 
contains all other eigenvalues. They correspond to eigenvectors that are constant on the orbits of the symmetric structure. 
These results  suggest that reducing the network by collapsing all nodes in each orbit onto each other leaves the spectrum of the adjacency invariant, except for removing the redundant eigenvalues.  

In the following we show that very similar results can be obtained for the Jacobian matrix of the food web model. When the redundant eigenvalues have a negative real part the resulting reduction leads to a lower-dimensional, but dynamically equivalent, system. 
To establish this reduction we consider the Jacobian as the adjacency matrix prescribing the weights and directions of the links in a (hypothetical) network. 

We note that for food webs the network prescribed by the Jacobian matrix is not identical to the network of predator-prey interactions. For instance, additional mutualistic
interactions may appear between different prey populations of a given predator \cite{Thilo2006}. However, if different populations
belong to an orbit in the network of predator-prey interactions, and if their generalized parameters are identical, then they also
form an orbit in the network defined by the Jacobian matrix. We therefore only have to show that the results of Ref.~\onlinecite{MacArthur2009} extend to such weighted and directed networks.

Following Ref.~\onlinecite{MacArthur2009} we argue that each symmetry generates an invariant subspace of the Jacobian matrix. 
Notably, this subspace contains only vectors that are zero outside the symmetry
and thus acts as the foundation for the set of eigenvectors that are localized on the symmetric structure.

We consider a subspace $V$ spanned by vectors $v$ that are zero except in the elements belonging to the symmetric structure.
The defining feature of the independent subspace is that $\matr{J}\vect{v}$ must lie within $V$.
This condition is met when $v$ obeys the ``zero-sum'' condition for each orbit of the symmetric structure, i.e. $\sum v_i =0$, where $i$  runs over the elements of the orbit.

To test  the localization  we first consider a node $k$ outside the symmetric structure. 
If this node is connected to any node within an orbit then symmetry implies that it must be connected to all nodes within the same orbit.
Further, all of these connections must carry the same weight (say, $b$). 
We can therefore write the terms of the $k$'th element of $\matr{J}\vect{v}$ arising from this orbit as $b\sum v_i =0$.  
This reasoning shows that the vector $v$ stays localized under application of $\matr{J}$.
 
Second, we have to show that $\matr{J}\vect{v}$ still meets the zero-sum condition, so that a vector remains localized under iterated application of the Jacobian. For this purpose let us consider the connections of an orbit $p$ with the nodes of an orbit  $q$ under the action of the Jacobian. 
One can always find a reordering of the variables such that all terms representing the links from nodes in $p$ towards nodes in $q$ appear in a single block in $\matr{J}$. In the following we refer to such blocks as \emph{orbit connections} and denote them by $\OrbC{qp}$.

We first show that the sum over the elements of $\OrbC{qp}\vect{v}$ belonging to nodes in $q$ is  zero.
More explicitly, we can write this sum as $\sum_i\left(\OrbC{qp}\vect{v}\right)_i$, where $i$ runs over the nodes in $q$ (over all elements of the vector  $\OrbC{qp}\vect{v}$). 
However, this is identical to first taking the sum over the elements of each column of $\OrbC{qp}$ and then multiplying by the vector $\vect{v}$. 
The column sums of $\OrbC{qp}$ correspond each to the sum over all links originating from one node in $p$ and are therefore identical by symmetry. Let this sum be $c$. We can then write $\sum_i\left(\OrbC{qp}\vect{v}\right)_i=c\sum_j v_j=0$, where $i$ runs over nodes in $q$ and $j$ runs over all nodes in $p$. 
In other words, this shows that the sum of the input that a given orbit receives from any other orbit under application of the Jacobian on a vector in $V$ sums to $0$.  
Consequently the zero-sums conditions remain fulfilled for $\matr{J}\vect{v}$ such that $v$ stays localized also under iterative application of the Jacobian matrix.

Because the two conditions, localization and zero-sums, are met by $\matr{J}\vect{v}$, $V$ is an independent subspace under $\matr{J}$, called the \emph{redundant} subspace of the symmetry. We note that the redundant subspace of a symmetry of $o$ orbits and of size $s$ has dimension $o(s-1)$.
Further, the compliment of the redundant subspace $W=V^\bot$ is also an independent subspace. The orthogonality of $W$ and $V$ implies that any vector $\vect{w}$ in $W$ is constant on nodes belonging to the same orbit. 

Intuitively speaking the redundant subspace $V$ corresponds to eigenmodes of the Jacobian that characterize perturbations affecting nodes in an orbit anti-symmetrically. Thus the effect on nodes outside the respective symmetric structure cancels, which explains the localization. The subspace $W$ correspond to eigenmodes of the Jacobian that characterize perturbations affecting nodes in the orbit identically. Thus the effect on nodes outside the symmetric structure does not cancel and couples the orbit to the rest of the system.

So far we have shown that the presence of a symmetric structure can be linked directly to the redundant eigenvalues in the spectrum of the Jacobian. These eigenvalues and the corresponding eigenvectors correspond to local eigenmodes governing the approach or departure of the system to or from the steady state. Therefore symmetric structures provide a rare example of mesoscale motifs that have exactly prescribed implications on the system-level dynamics. 

For computing the redundant eigenvalues, one needs to diagonalize the Jacobian only in the subspace $V$. The difficulty involved thus depends on the complexity of the symmetric structure, but not on the complexity of the embedding system.      

One can now imagine a mapping which collapses all nodes in a given orbit onto each other such that each orbit is replaced with a single substitute node. Intuitively, this mapping should remove the redundant eigenvalues from the spectrum of $\matr{J}$ while leaving all other eigenvalues unchanged.

For confirmation we establish $\matr{J}$ for vectors inside $W$ in terms of new variables $\vect{w}'^{p}$, each of which replaces all the (identical) variables of an orbit $p$. We recall that this substitution is possible because vectors in $W$ are constant on each orbit. 

More explicitly, the elements of the vector $\matr{J}\vect{w}$ corresponding to nodes of given orbit $q$ are given by $\sum_p \OrbC{qp}\vect{1} w'^{p} $, where $p$ runs over all orbits, $\OrbC{qp}$ denotes the orbit connection from $p$ to $q$, and $\vect{1}$ is a vector that contains the entry $1$ for each node in $q$.
Because each of the row-sums of $\OrbC{qp}$ represents the sum over the incoming links of one node in $q$ from nodes in $p$, the entries of $\OrbC{qp}\vect{1}$ representing these row-sums are identical. We can therefore consistently replace the entries in $\sum_p \OrbC{qp}\vect{1} w'^{p}$ by $w'^q$.
Hence the connection from the variables ${w'}^{p}$ to  $w'^{p}$ is given by the row-sum of $\OrbC{qp}$.
This defines the \emph{reduced Jacobian matrix} $\matr{J}'$ which contains at each position this connection from $w'^{q}$ to $w'^{p}$. Furthermore it defines the reduced (hypothetical) network of $\matr{J}'$ which contains one node for each replacement variable ($w'^{p}$ instead of the orbit $p$), confirming the intuitive assumption.

These results can be summarized in a simple algorithm which reduces the network of the Jacobian  to the network of the reduced Jacobian.
Starting with the graph of the Jacobian  $\matr{J}$, we eliminate all nodes except one for each orbit. Links that pointed towards an eliminated node are removed and links that originated from an eliminated node are rewired to the remaining node from the same orbit. 
Finally, we simplify the network by replacing all links originating and ending in the same nodes by a single new link. The weight of this replacement link is the sum over the replaced ones.
Repeating this procedure if new symmetries arise from the reduction we ultimately obtain the so-called quotient graph corresponding to the reduced Jacobian  which contains no more symmetries.

The spectrum of the reduced Jacobian matrix $\matr{J}'$ describes the dynamics of the original system, except for the localized redundant part. However, if the redundant eigenvalues of $\matr{J}$ have negative real part, they do not change the dynamics qualitatively and analysing $\matr{J}'$ is equivalent to analysing $\matr{J}$. 

Once the Jacobian has been equivalently reduced, one can try to construct a physical representation of the reduced system.  We illustrate this process for food webs in the next section.

\section{Application to Food Webs\label{sec:Results}}
The results in \secref{sec:Seperation} imply that the nodes of each symmetry in a predator-prey network carry localized dynamical modes inside the associated redundant subspace of the Jacobian matrix. One can imagine that reducing the food web model by collapsing the symmetric nodes removes the modes associated with the symmetry but leaves the remaining dynamical modes intact. Thus  this reduction leads to an equivalent dynamical system, if the removed modes are stationary and stable, i.e. when the redundant eigenvalues have negative real parts.
In this section we confirm this intuition by analysing different symmetric structures inside simple example food webs.

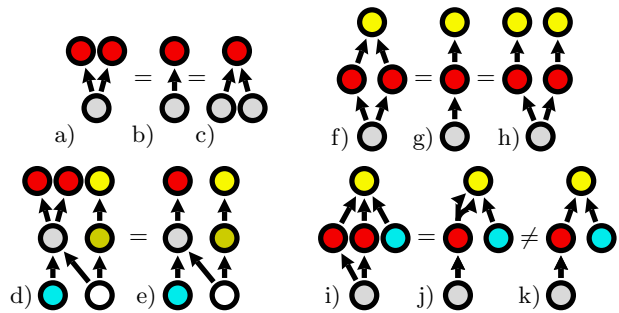
\begin{figure}[tb]
\centering
 \psset{yunit=0.76,xunit=0.83,fillstyle=solid,linewidth=0.07,linecolor=black,fillcolor=white, nodesep=0.03, arrowinset=0.1, arrowlength=1}
\def\radius{0.2}
 \def\CompEx{
\begin{pspicture}(-0.5,-1.75)(2.25,0.25)

\rput[b](-0.5,-1.7){a)}

\cnode[fillcolor=orb1](-0.25,0){\radius}{CP1}
\cnode[fillcolor=orb1](0.25,0){\radius}{CP3}
\cnode[fillcolor=attorb](0,-1){\radius}{CJ1}
\ncline{->}{CJ1}{CP1}
\ncline{->}{CJ1}{CP3}

\rput[c](0.75,-0.5){$=$}

\cnode[fillcolor=orb1](1.25,0){\radius}{CP}
\cnode[fillcolor=attorb](1.25,-1){\radius}{CJ}
\ncline{->}{CJ}{CP}

\rput[b](0.75,-1.7){b)}

\rput[c](1.6125,-0.5){$=$}

\cnode[fillcolor=attorb](2,-1){\radius}{PPP1}
\cnode[fillcolor=attorb](2.5,-1){\radius}{PPP3}
\cnode[fillcolor=orb1](2.25,0){\radius}{PPJ}
\ncline{->}{PPP1}{PPJ}
\ncline{->}{PPP3}{PPJ}

\rput[b](1.75,-1.7){c)}
\end{pspicture}
}

\def\LargeCompEx{
\begin{pspicture}(-0.5,-1.25)(3,1.25)

\rput[b](-0.5,-1.25){d)}

\cnode[fillcolor=orb1](-0.25,1){\radius}{LCT1}
\cnode[fillcolor=orb1](0.25,1){\radius}{LCT2}
\cnode[fillcolor=subo1](0.75,1){\radius}{LCT3}
\cnode[fillcolor=attorb](0,0){\radius}{LCM1}
\cnode[fillcolor=subo2](0.75,0){\radius}{LCM2}
\cnode[fillcolor=orb2](0,-1){\radius}{LCB1}
\cnode[fillcolor=white](0.75,-1){\radius}{LCB2}

\ncline{->}{LCM1}{LCT1}
\ncline{->}{LCM1}{LCT2}
\ncline{->}{LCM2}{LCT3}
\ncline{->}{LCB1}{LCM1}
\ncline{->}{LCB2}{LCM1}
\ncline{->}{LCB2}{LCM2}

\rput[c](1.375,0){$=$}

\rput[b](1.5,-1.25){e)}

\cnode[fillcolor=orb1](2,1){\radius}{RLCT}
\cnode[fillcolor=subo1](2.75,1){\radius}{RLCT3}
\cnode[fillcolor=attorb](2,0){\radius}{RLCM1}
\cnode[fillcolor=subo2](2.75,0){\radius}{RLCM2}
\cnode[fillcolor=orb2](2,-1){\radius}{RLCB1}
\cnode[fillcolor=white](2.75,-1){\radius}{RLCB2}

\ncline{->}{RLCM1}{RLCT}
\ncline{->}{RLCM2}{RLCT3}
\ncline{->}{RLCB1}{RLCM1}
\ncline{->}{RLCB2}{RLCM1}
\ncline{->}{RLCB2}{RLCM2}

\end{pspicture}
}

\def\Jpi{
\begin{pspicture}(-0.75,-1.25)(3.4,1.25)

\rput[b](-0.325,-1.25){f)}

\cnode[fillcolor=subo1](0.175,1){\radius}{I}
\cnode[fillcolor=orb1](-0.15,0){\radius}{P1}
\cnode[fillcolor=orb1](0.5,0){\radius}{P3}
\cnode[fillcolor=attorb](0.175,-1){\radius}{J}

\ncline{->}{J}{P1}
\ncline{->}{J}{P3}
\ncline{->}{P1}{I}
\ncline{->}{P3}{I}

\rput[c](1,0){$=$}

\rput[b](1,-1.25){g)}

\cnode[fillcolor=subo1](1.5,1){\radius}{IR}
\cnode[fillcolor=orb1](1.5,0){\radius}{QR}
\cnode[fillcolor=attorb](1.5,-1){\radius}{JR}

\ncline{->}{JR}{QR}
\ncline{->}{QR}{IR}

\rput[c](2,0){$=$}

\rput[b](2.375,-1.25){h)}
\cnode[fillcolor=subo1](2.5,1){\radius}{IM1}
\cnode[fillcolor=subo1](3.15,1){\radius}{IM2}
\cnode[fillcolor=orb1](2.5,0){\radius}{QM1}
\cnode[fillcolor=orb1](3.15,0){\radius}{QM2}
\cnode[fillcolor=attorb](2.875,-1){\radius}{JM}
\ncline{->}{QM1}{IM1}
\ncline{->}{QM2}{IM2}
\ncline{->}{JM}{QM1}
\ncline{->}{JM}{QM2}
\end{pspicture}}

\def\Jri{
\begin{pspicture}(-0.75,-1.25)(4.05,1.25)

\rput[b](-0.5,-1.25){i)}
\cnode[fillcolor=subo1](0,1){\radius}{I}
\cnode[fillcolor=orb1](-0.5,0){\radius}{P1}
\cnode[fillcolor=orb1](0,0){\radius}{P2}
\cnode[fillcolor=orb2](0.5,0){\radius}{P3}
\cnode[fillcolor=attorb](0,-1){\radius}{J2}
\ncline{->}{J2}{P1}
\ncline{->}{J2}{P2}
\ncline{->}{P1}{I}
\ncline{->}{P2}{I}
\ncline{->}{P3}{I}

\rput[c](1,0){$=$}

\rput[b](1,-1.25){j)}
\cnode[fillcolor=subo1](1.825,1){\radius}{RI}
\cnode[fillcolor=orb1](1.5,0){\radius}{RP1}
\cnode[fillcolor=orb2](2.15,0){\radius}{RP3}
\cnode[fillcolor=attorb](1.5,-1){\radius}{RJ1}
\ncline{->}{RJ1}{RP1}
\ncline{->}{RJ1}{RP1}
\nccurve[angleA=90,angleB=225]{->}{RP1}{RI}
\nccurve[angleA=90,angleB=255]{->}{RP1}{RI}
\ncline{->}{RP3}{RI}

\rput[c](2.65,0){$\neq$}

\rput[b](2.65,-1.25){k)}
\cnode[fillcolor=subo1](3.475,1){\radius}{RIS}
\cnode[fillcolor=orb1](3.15,0){\radius}{RPS1}
\cnode[fillcolor=orb2](3.8,0){\radius}{RPS3}
\cnode[fillcolor=attorb](3.15,-1){\radius}{RJS1}
\ncline{->}{RJS1}{RPS1}
\ncline{->}{RJS1}{RPS1}
\ncline{->}{RPS1}{RIS}
\ncline{->}{RPS3}{RIS}
\end{pspicture}
}

\def\JriAlternative{
\begin{pspicture}(-0.75,-1.25)(4.05,1.25)

\rput[b](-0.5,-1.25){i)}
\cnode[fillcolor=subo1](0.5,1){\radius}{I}
\cnode[fillcolor=orb1](-0.5,0){\radius}{P1}
\cnode[fillcolor=orb1](0,0){\radius}{P2}
\cnode[fillcolor=orb2](0.5,0){\radius}{P3}
\cnode[fillcolor=attorb](0.0,-1){\radius}{J}
\ncline{->}{J}{P1}
\ncline{->}{J}{P2}
\ncline{->}{J}{P3}
\ncline{->}{P3}{I}

\rput[c](1,0){$=$}

\rput[b](1.325,-1.25){j)}
\cnode[fillcolor=subo1](2.15,1){\radius}{RI}
\cnode[fillcolor=orb1](1.5,0){\radius}{RP1}
\cnode[fillcolor=orb2](2.15,0){\radius}{RP3}
\cnode[fillcolor=attorb](1.825,-1){\radius}{RJ}
\ncline{->}{RP3}{RI}
\nccurve[angleB=280,angleA=105]{->}{RJ}{RP1}
\nccurve[angleB=280,angleA=135]{->}{RJ}{RP1}
\ncline{->}{RJ}{RP3}

\rput[c](2.65,0){$\neq$}

\rput[b](2.975,-1.25){k)}
\cnode[fillcolor=subo1](3.8,1){\radius}{RIS}
\cnode[fillcolor=orb1](3.15,0){\radius}{RPS1}
\cnode[fillcolor=orb2](3.8,0){\radius}{RPS3}
\cnode[fillcolor=attorb](3.475,-1){\radius}{RJS}
\ncline{->}{RJS}{RPS3}
\ncline{->}{RJS}{RPS1}
\ncline{->}{RPS3}{RIS}
\end{pspicture}
}
\CompEx \hspace{20pt} \Jpi \\
\vspace{5pt}
\LargeCompEx \hspace{20pt} \Jri
\caption{Examples for food webs that differ only by dynamical modes localized inside their mesoscale symmetries. Different colours denote different orbits. Two systems are marked as dynamically equivalent (``$=$'') if the spectra of their Jacobian matrices are identical up to the eigenmodes characterizing the mesoscale effects inside the symmetries.
If nodes of a symmetry have a common prey or predator with a node outside the symmetry, as e.g. in i), the generalized parameters of the common prey or predator in the reduced network are weighted which is shown in j) as a double-headed link. See text for more details.\label{fig:ReducWebs}}
\end{figure}

We start by considering the so-called competitive-exclusion motif shown in \figref{fig:ReducWebs}a). In this motif two top-predators (populations $1,2$) feed on the same prey (population $3$).
Using Ref.~\onlinecite{ThiloLars2009} the stability of any given steady state is governed by a Jacobian of the form 
\eq{\matr{J}=
\begin{pmatrix} J_{11}	&	0 	&  J_{13}\\
		    0     	&	J_{22}	&  J_{23}\\
		    J_{31}	&	J_{32}	&  J_{33}
\end{pmatrix}
\label{eq:CEJ}.}

The diagonal terms reflect for each species the dependence on its own abundance of mortality, of primary production, and of the predatory losses and gains. The off-diagonal terms reflect the dependence of the predator-prey interactions on either the predator or the prey abundance.

This motif contains a symmetry if the generalized parameters of population $1$ and $2$ are identical (such that $J_{13}$=$J_{23}$, $J_{11}$=$J_{22}$,$J_{31}$=$J_{32}$) and simple diagonalization yields the three eigenvalues $\lambda_{3}=J_{11}$ and $\lambda_{1,2} =\ttfrac{1}{2}(J_{11}+J_{33}\pm\sqrt{\Delta})$, where $\Delta=(J_{11}-J_{33})^2+4J_{31}J_{31}$.

Following the procedure outlined in the previous section, we can write the Jacobian of a smaller system
\eq{
\tilde{\matr{J}}=
\begin{pmatrix} J_{11} &  	J_{13}	\\
		2J_{31}	&   	J_{33}
\end{pmatrix},
\label{eq:ChainJ}} 
that exhibits qualitatively similar dynamics.
Straightforward diagonalization confirms that this reduced matrix $\tilde{\matr{J}}$ retains Eigenvalues $\lambda_{1,2}$, while the eigenvalue $\lambda_3$ corresponding to the mode associated with the mesoscale symmetry has disappeared.

We note that the reduced matrix $\tilde{\matr{J}}$ can be interpreted as the Jacobian of a predator-prey system. This is not trivial because the terms such as $J_{13}$ and $J_{31}$ are not independent, but can be expressed as functions of the same set of generalized parameters. Using the detailed representation of these terms \cite{Thilo2005,ThiloFeudel2008} we verified (c.f. \ref{sec:AppendixCE}) that $\tilde{\matr{J}}$ corresponds to a predator-prey system, in which the prey is described by the same parameters as species $1$ in the original system and the predator by the parameters of species $3$. It is remarkable that the factor 2 in $\tilde{\matr{J}}$ that arises from the reduction is indeed needed to make this direct correspondence possible. 

The dynamics of the reduced (predator-prey) system is equivalent to the original (competitive-exclusion) system if the dynamics in the direction of the redundant eigenvector is stable, i.e. if the eigenvalue $\lambda_3$ that disappears in the reduction has a negative real part. For the competitive-exclusion motif this is typically the case when the predators suffer from super-linear mortality \cite{GrossEdwardsFeudel2009}.

While we have so far focused on the three node web for illustration, we emphasize that the same reduction can be applied in food webs of arbitrary size. In particular, using the results from the previous section, it is not necessary to compute the complete spectrum of the Jacobian to apply the reduction. For instance, consider the competitive-exclusion motif inside the complicated food web in \figref{fig:ReducWebs}d) for which writing all the Jacobian matrix eigenvalues is very complicated. Even in this case the results from \secref{sec:Seperation} guarantee that the the competitive exclusion motif that appears in the web carries the redundant eigenvector associated with the eigenvalue $\lambda_{3}$ found before. Removing the associated dynamical modes leads to the Jacobian matrix of the reduced food web in \figref{fig:ReducWebs}e), which is dynamically equivalent if $\lambda_{3}<0$.

As a second example we consider the so-called apparent-competition motif shown in \figref{fig:ReducWebs}c), i.e. one predator (population $1$) feeding on two prey species (populations $2,3$).
The Jacobian matrix is 
\eq{\matr{J}=
\begin{pmatrix} J_{11}	&	J_{12} 	&  J_{12}	 \\
		    J_{21}     	&	J_{22}	&  J_{23}\\
		    J_{21}	&	J_{23}	&  J_{22}
\end{pmatrix},
\label{eq:PPJ}} where we have assumed that the prey species are described by identical ecological parameters.
In contrast to the Jacobian of the competitive-exclusion motif this matrix contains the elements denoted by $J_{23}$ that couple the symmetric (prey) species. These so-called apparent competition terms arise because feeding on one prey species leads to a saturation of the predator that benefits both prey species. By contrast the two predators in the competitive-exclusion motif can generally be assumed to feed on the prey independently of each other.

The dynamical mode localized on the two prey species is characterized by the eigenvalue $\lambda_3=J_{22}-J_{23}$ and the corresponding eigenvector  $(0,1,-1)^T$. 
Therefore if $J_22<J_{23}$, which is generally the case in food webs, then removing the dynamical mode of the symmetry leads to a dynamically equivalent system.

Following the procedure outlined in \secref{sec:Seperation}, we can write the Jacobian of  the reduced system as
\eq{\tilde{\matr{J}}=
\begin{pmatrix} J_{11} & 2 J_{12} \\
		J_{21}	& J_{22}+J_{23}
\end{pmatrix}
\label{eq:PPReducedJ},} which retains the eigenvalues of $\matr{J}$ except for $\lambda_3$.
Comparing $\tilde{\matr{J}}$ to the Jacobian matrix of the system in \figref{fig:ReducWebs}b) we find that $\tilde{\matr{J}}$ describes a predator-prey system with identical generalized parameters (c.f \ref{sec:AppendixPP}).

The remaining examples in \figref{fig:ReducWebs} can be treated along the same lines as above.
Let us therefore comment only briefly on two new features that appear in these examples.
First, when aggregating symmetric species that share a common predator with other species that are not part of the same orbit then care has to be taken that the aggregate species enters with correct weight.  
For instance in \figref{fig:ReducWebs}i), if we assume that the top predator receives two third of its biomass from the two symmetric species. Then in the reduced system the species replacing the orbit has twice the weight in comparison to the non-symmetric prey on the right.
    
Second, we note that symmetries consisting of more than one orbit may exhibit more complicated localized dynamics.
For instance in the food web shown in \figref{fig:ReducWebs}h) the symmetric structure consists of two interconnected orbits.
Diagonalization yields the redundant eigenvalues
$\lambda_{1,2} =\ttfrac{1}{2}(J_{11}+J_{33}\pm\sqrt{\Delta})$ with $\Delta=(J_{11}-J_{33})^2+4J_{31}J_{13}$,  where we used the indices $1$ and $3$ to denote a top-predator and its prey, respectively.
The redundant eigenvalues can thus form a complex-conjugate eigenvalue pair, potentially leading to oscillatory instabilities.

We emphasize that symmetries present in the steady state do not generally persist when the system departs from the this state. Dynamics that is generated in a smymmetric mesoscale motif can therefore spread and impact the entire network. For instance in the example discussed above the oscillatory instability can be inferred solely from the analysis of the mesoscale motif, wheras the resulting oscillations may affect the entire network.

In summary every mesoscale symmetry is associated with one or more dynamical modes that originate in the symmetric motif. If the dynamical modes are stable, then they can be removed from the system  without changing the remaining dynamics by replacing each orbit of the symmetry by a single node.
For this reduced system the Jacobian can be obtained according to the rules described in \secref{sec:Seperation}. The reduced Jacobian can in general be interpreted as describing a smaller food web, however, potentially with altered parameter values.
For the simple cases considered in Ref.~\onlinecite{ThiloFeudel2008} the parameters of the original predator-prey network and those of the reduced network are so closely related that the dynamically equivalent systems could be spotted empirically. However, this is  much harder for complicated topologies, as e.g. in \figref{fig:ReducWebs}d), and for more elaborate models. Even in such complicated cases the aggregated Jacobian can be computed relatively straight forwardly with the procedure proposed here. Notably the complexity of this reduction increases with the size of the symmetric structure, but is independent of the size of the embedding network.

\section{Conclusion}

In this paper we showed that a previously observed equivalence in the dynamics of food webs \cite{ThiloFeudel2008} is a consequence of localized dynamical modes originating from on  mesoscale symmetries.

We extended a recent result \cite{MacArthur2009} to show that certain dynamical modes localize on symmetric structures in food webs. In cases where such mesoscale structures are present these dynamical modes are independent from the network outside the symmetry.
If the localized modes are stable, then aggregating the symmetric structures in a food web as described in  \secref{sec:Seperation} removes them while leaving the remaining dynamics intact, leading to a smaller but dynamically equivalent system.

In principle the findings of this paper can be used to formulate reduced models for certain food webs. In practice this application of the results will be of little importance as removing symmetries leads, at best, to a relatively mild aggregation. 

However, from the point of view of physics, symmetric motifs in networks, provide an example of mesoscale structures that have distinct and exact dynamical implications for the dynamics of the network as a whole.  
These implications can be explored even in very large networks because computation of the corresponding localized eigenvector and eigenvalue do not require the computation of the complete spectrum of the Jacobian.

\begin{appendix}
\section{Reduction of the Jacobian matrix of the Competitive-Exclusion Motif \label{sec:AppendixCE}}
In this appendix we show the explicit reduction of the Jacobian matrix for the competitive exclusion motif in \figref{fig:ReducWebs}a) in terms of generalized modeling parameters of the model presented in Ref.~\onlinecite{Thilo2005,ThiloLars2009}. Furthermore we show that the reduced Jacobian matrix $\tilde{\matr{J}}$ is indeed identical to the Jacobian matrix of the simple predator-prey system in \figref{fig:ReducWebs}b).
In terms of generalized modeling parameters \eqref{eq:CEJ} becomes
\begin{align*}\matr{J}= \begin{pmatrix}
               \alpha_1 &	0 &	0	\\
		0	&\alpha_1	&	0\\
		0	&	0	&\alpha_3\\
              \end{pmatrix}
	      \begin{pmatrix}
		\Psi_1-\mu_1	&	0	&	\gamma_1	\\
		0		&\Psi_1-\mu_1	&	\gamma_1	\\ 
	        -\frac{1}{2}\Psi_1&	-\frac{1}{2}\Psi_1	&\Phi_3-\gamma_1\\
	      \end{pmatrix},\nonumber
\end{align*}
where $\alpha_i$ is the biomass turnover rate of species $i$,
$\Psi_1$ is the sensititivity of predation to the abundance of both predators, $\mu_1$ is the sensitivity of mortality to predator abundance of both predators, $\Phi_3$ is the sensitivity of primary production to the abundance of prey and $\gamma_1$ is the sensitivity of predation for both predators to the abundance of prey. The factors of $\frac{1}{2}$ in the last row arise because each of the species $1$ and $2$ is responsible for half of the biomass loss of species $3$ also their effect on species $3$ is weighed by this importance.

Using the results in \secref{sec:Seperation}, the reduced Jacobian matrix (c.f. \eqref{eq:ChainJ}) is 
\begin{align*}\tilde{\matr{J}}= \begin{pmatrix}
               \alpha_1 &	0\\
		0	&\alpha_3\\
		           \end{pmatrix}
	      \begin{pmatrix}
		\Psi_1-\mu_1	&	\gamma_1	\\
	        2\left(-\frac{1}{2}\Psi_1\right)& 	\Phi_3-\gamma_1 \\
	      \end{pmatrix}.
\end{align*}
This is identical to the Jacobian matrix for a predator prey-system in which the predator is described by the generalized parameters $\alpha_1,\Psi_1,\gamma_1,\mu_1$ and the prey species by $\Phi_3$.

\section{Reduction of the Jacobian matrix of the Apparent-Competition Motif \label{sec:AppendixPP}}
In terms of generalized modeling parameters, the Jacobian matrix in \eqref{eq:PPJ} of the  apparent-competition motif in \figref{fig:ReducWebs}c) is
\begin{align*}\matr{J}= \begin{pmatrix}
               \alpha_1 &	0 &	0	\\
		0	&\alpha_2	&	0\\
		0	&	0	&\alpha_2\\
              \end{pmatrix}
      \begin{pmatrix}
	\Psi_1-\mu_1	&\frac{1}{2}\gamma_1		&\frac{1}{2}\gamma_1	    \\
	-\Psi_1		& \Phi_2 -\frac{\gamma_1+1}{2}	& -\frac{\gamma_1-1}{2}	    \\ 
	-\Psi_1		& -\frac{\gamma_1-1}{2}		& \Phi_2 -\frac{\gamma_1+1}{2}\\
      \end{pmatrix},\nonumber
\end{align*}
where the parameters are the same as in \ref{sec:AppendixCE}. The factors of $\frac{1}{2}$ in the matrix arise because each of the two prey species is responsible for half of the biomass gain of species $1$. In comparison to the Jacobian of the competitive-exclusion motif, the matrix contains the elements $-\frac{\gamma_1-1}{2}$ that couple the symmetric (prey) species. These are the apparent competition terms arising because feeding on one prey species leads to a saturation of the predator that benefits both prey species.

Using the results in \secref{sec:Seperation}, the reduced Jacobian matrix is 
\begin{align*}\tilde{\matr{J}}= \begin{pmatrix}
               \alpha_1 &	0\\
		0	&\alpha_2\\
		           \end{pmatrix}
	      \begin{pmatrix}
		\Psi_1-\mu_1	&	2\frac{1}{2}\gamma_1	\\
	        -\Psi_1& 	 \Phi_2 -\gamma_1 \\
	      \end{pmatrix}.
\end{align*}
Therefore $\tilde{\matr{J}}$ is also identical to the Jacobian matrix of the simple predator-prey system in which the predator is also described by the generalized parameters $\alpha_1,\Psi_1,\gamma_1,\mu_1$ and the prey species by $\Phi_3$.
\end{appendix}

\addcontentsline{toc}{section}{Bibliography}

\end{document}